\documentclass[aps,prd,groupedaddress,nofootinbib]{revtex4}

\usepackage{graphicx}
\usepackage{amssymb,amsmath}
\usepackage[usenames,dvipsnames]{color}

\newcommand{\be}{\begin{equation}}
\newcommand{\ee}{\end{equation}}
\newcommand{\bea}{\begin{eqnarray}}
\newcommand{\eea}{\end{eqnarray}}	

\newcommand{\ba}{\begin{array}}
\newcommand{\ea}{\end{array}}

\definecolor{MyDarkGray}{RGB}{140,140,140}

\begin{document}

\title{Inflation induced by Gravitino Condensation in Supergravity\\
~\\
~\\}



\author{John Ellis and Nick E. Mavromatos}
\affiliation{~\\Theoretical Particle Physics and Cosmology Group, Department of Physics, King's College London, Strand, London WC2R 2LS, UK; \\
Theory Division, Physics Department, CERN, CH-1211  Geneva 23, Switzerland}


\begin{abstract} 
\vspace{1cm}
We discuss the emergence of an inflationary phase in supergravity with the super-Higgs effect 
due to dynamical spontaneous breaking of supersymmetry, in which the r\^ole of the inflaton is played by the gravitino condensate. 
Realistic models compatible with the Planck satellite CMB data are found in conformal supergravity scenarios with dynamical 
gravitino masses that are small compared to the Planck mass, as could be induced by a non-trivial vacuum expectation 
value of the dilaton superfield of appropriate magnitude.
\vspace{1cm}
\begin{center}
KCL-PH-TH/2013-24,
LCTS/2013-15,
CERN-PH-TH/2013-184
\end{center}

\end{abstract}

\maketitle

\section{Introduction}

Inflation~\cite{guth}, especially the version with a scalar field (inflaton) rolling slowly towards a non-trivial global minimum of its 
potential~\cite{linde,ast}, is a successful framework for explaining general features of cosmology and notably the data on the 
Cosmic Microwave Background (CMB) from the WMAP~\cite{wmap} and Planck satellites~\cite{Planck}. 
Although these measurements have excluded most of the monomial potentials for single-field inflation~\cite{encyclopaedia}, 
many models survive, including Starobinski's $R + R^2$ model~\cite{starobinski} and Higgs inflation and its 
variants~\cite{shaposhnikov,higgsdil}. 

It was suggested long ago that inflation cries out for spontaneously-broken supersymmetry~\cite{cries}, since a very flat 
inflationary potential would be more natural than in non-supersymmetric models. In a recent work~\cite{croon} 
we updated phenomenological studies of non-monomial inflationary potentials~\cite{encyclopaedia} within supersymmetry, 
showing that a toy flat-space supersymmetric Wess-Zumino model with a single chiral scalar superfield gives rise to an
inflaton potential that is comfortably consistent with the Planck data.

The inclusion of supergravity effects is desirable if not essential for realistic supersymmetric models of inflation, 
and the Starobinski-type inflationary potential has recently been shown~\cite{olive} to emerge from certain no-scale supergravity models.  
Several other interesting works on inflationary scenarios in supergravity models have appeared recently~\cite{sugrainfl}, 
reviving interest in the cosmological relevance of such models, and there have also been interesting new 
non-supersymmetric approaches~\cite{nonSUSY}. 

Complementary to the above approaches is the work of~\cite{alvarez}, where a minimal supergravity inflation scenario is  realized in the context of broken global supersymmetry, and the inflationary phase is connected with a renormalization-group 
flow from the Ultraviolet (UV) to the Infrared (IR) of a constrained chiral scalar superfield appearing~\cite{komar} in the broken 
superconformal current Ferrara-Zumino~\cite{ferrara} multiplet in theories with F-type breaking of supersymmetry. 
The IR limit of this chiral superfield constitutes the well-known Volkov-Akulov~\cite{va} Goldstino supermultiplet, 
whose spin-1/2 component is the Goldstino majorana fermion that appears in spontaneously-broken global supersymmetry.
In this scenario, the inflaton is identified with the scalar component of the chiral superfield in the UV limit.
Upon appropriate embedding in supergravity, its K\"ahler potential may be chosen so as to yield successful
small-field inflation in agreement with the Planck data~\cite{Planck}. It was also shown in this framework~\cite{alvarez}
that there is a relation between the scale of global supersymmetry breaking  and the amount of non-Gaussian 
fluctuations generated by the inflaton field.

In the present work we pursue another avenue for producing inflation in supergravity models, 
which may be realized in models in which local supersymmetry is broken dynamically. 
Such scenarios differ from those examined in~\cite{alvarez} in that in our approach it is the gravitino 
condensate field that plays the r\^ole of the inflaton via its one-loop effective potential. 
Nevertheless, as we discuss in detail, the presence of a super-Higgs effect~\cite{deser,cremmer},
in which global supersymmetry is broken at a scale $\sqrt{f}$ leading to a non-linear realization of supersymmetry
with a Volkov-Akulov Goldstino, is  essential in providing the scale of inflation, 
since the Hubble parameter is related to $\sqrt{f}$. However, unlike the situation in~\cite{alvarez}, 
knowledge of the potential of the scalar supermultiplet of the Ferrara-Zumino current conformal anomaly~\cite{ferrara} 
in the UV is not necessary. In our scenario, small-field inflation is realized in the IR limit
via the one-loop effective potential of the gravitino condensate field that is formed as a result of the four-fermion 
gravitino interactions in the torsion part of the minimal ${\mathcal N}=1$ supergravity action~\cite{VanNieuwenhuizen:1981ae}~\footnote{For early work on gravitino condensates and a possible application to inflation, 
within extended (superstring-inspired higher-dimensional) supergravity models,
see~\cite{duff,Oh}.}. Our analysis of the one-loop effective potential is performed in Minkowski space-time, 
as in~\cite{smith,smith2}~\footnote{A full analysis of the one-loop effective action of ${\mathcal N}=1$ supergravity
has been performed in~\cite{alexandre}, taking quantum gravity metric fluctuations into account as in~\cite{fradkin}, 
with similar results for the shape of the effective potential and its link to inflation as our flat-space analysis here. 
That work also counters the objections of \cite{odintsov} to the work of \cite{smith,smith2}, 
by finding cases in which there are no imaginary parts in the effective action coming from quantum metric fluctuations.}. 
We ignore consistently the quantum (super)gravity corrections by working~\cite{emdyno} 
with a conformal supergravity action~\cite{ferrara2} in which 
the dilaton is stabilized by a suitable potential, which we do not specify in our analysis here~\footnote{In our case we shall deal with single-field inflation, since 
the gravitino condensate is a single real scalar field. Hence non-Gaussianities are not present when the dynamics of the dilaton is ignored, as assumed here.}.
We assume that it determines a vacuum expectation value (v.e.v.) that fixes the coupling of the four-gravitino torsion self-interactions, 
${\tilde \kappa} = e^{-\langle \varphi \rangle} \, \kappa$ to be stronger
than the standard gravitational coupling, $\kappa $, in the Einstein-frame formalism of the ${\mathcal N}=1$ conformal supergravity. 
In this way, quantum-gravity corrections can be safely ignored when constructing the potential
that is used in our study of inflation. 

The shape of the potential depends on a cut-off scale, and a flat potential near the origin that leads to small-field inflation
can be obtained for values of the cut-off that are low relative to the coupling ${\tilde \kappa}$. Agreement with the Planck data~\cite{Planck} 
can be obtained for a suitable value of the cut-off relative to a transmutation mass scale. 
This yields a minimal inflationary scenario in the simplest (conformal) supergravity model, 
which does not require knowledge of the K\"ahler potential, unlike other models in the literature. 
Moreover, as already mentioned, our scenario differs from that of~\cite{alvarez} in that it occurs in the IR
with the r\^ole of the inflaton being played by the gravitino condensate, not the UV scalar supermultiplet 
that contains the Goldstino~\footnote{Of course, as the latter requires a redefinition of the gravitino
when the super-Higgs effect is in operation~\cite{deser}, eliminating the Goldstino from the physical spectrum 
and making the gravitino massive, the condensate contains contributions from the Goldstino, but the underlying physics is different.}.
The inflationary era in our scenario coincides with a flow of the gravitino condensate field
towards its non-trivial minimum at the end of inflation. 
The space-time at the end of inflation is flat Minkowski in our toy example of simple (conformal) supergravity 
without coupling to matter~\footnote{The coupling to matter leads to decays of the gravitino
and subsequent reheating, but such issues are beyond the scope of this article.}. 
We stress that one feature of such a conformal simple supergravity model is that 
the constraint imposed by the CMB data~\cite{encyclopaedia} on the scale of the potential relative
to the slow-roll parameter $\epsilon$ is satisfied for relatively large conformal gravitational coupling:
${\tilde \kappa} \gg \kappa$, which is an essential feature of our approach.
 
The structure of the article is as follows: in Section~\ref{sec:sh} we review the super-Higgs effect 
and the breaking of local supersymmetry, and the resulting effective infrared supergravity Lagrangian 
with a cosmological constant. In Section~\ref{sec:oneloop} we construct and study the structure of the 
one-loop effective action in flat Minkowski space-time of the gravitino condensate in conformal 
${\mathcal N}=1 $ supergravity models. The details of our inflationary scenario
based on the corresponding one-loop effective potential are discussed in Section~\ref{sec:infl},
followed by a comparison with current data. Finally, Section~\ref{sec:concl} summarizes our conclusions and the outlook.

\section{The Super-Higgs Effect and the Dynamical Breaking of Conformal Supergravity \label{sec:sh}}

We start by reviewing the situation where global supersymmetry is broken by an appropriate F-term due to some chiral superfield 
acquiring a vacuum expectation value  :
 \begin{equation}\label{fterm}
		\langle F \rangle \; = \; f \; \ne \; 0 ~.
	\end{equation}
The corresponding Goldstone field is a fermion called the Goldstino, a Majorana fermion field $\lambda (x)$ with spin 1/2, 
whose low-energy interactions are described by a Volkov-Akulov-type Lagrangian~\cite{va} that realizes global supersymmetry
non-linearly:
\begin{equation}\label{val}
 \mathcal{L}_\lambda \; = \; -(f^2){\det}\left(\delta^\mu_\nu + i \frac{1}{2f^2}\overline{\lambda} \gamma^\nu \partial_\mu \lambda \right) 
\end{equation}
The infinitesimal parameter of the non-linear realization of global supersymmetry is $\alpha$:
\begin{equation}\label{trnl}
\delta \lambda \; = \;  f\,\alpha + i \frac{1}{f} \overline{\alpha} \gamma^\mu \lambda \partial_\mu \lambda~.
\end{equation}
The coupling of the Goldstino to supergravity generates a mass for the gravitino through the absorption of
the Goldstino, via the super-Higgs effect envisaged in~\cite{deser,cremmer},
and we now review some aspects that are relevant for our discussion. 

We consider $\mathcal{N}=1$ supergravity theory in four space-time dimensions in the second-order 
formalism~\cite{VanNieuwenhuizen:1981ae}:
\begin{eqnarray}\label{msugra}
e^{-1}  \mathcal{L}   \; = \;   -\frac{1}{2\kappa^2} R(e)  - \frac{1}{2} \epsilon^{\mu\nu\rho\sigma} \overline{\psi }_\mu  \gamma_5 \gamma_\nu D_\rho {\psi}_\sigma 
 -  \frac{11 \kappa^2}{16}  \left[ (\overline{\psi}_\mu  {\psi}^\mu )^2  - (\overline{\psi}_\mu  \gamma_5 {\psi}^\mu )^2 \right]
+ \frac{33}{64} \kappa^2  \, \left(\overline{\psi}^\mu \gamma_5  \gamma_\nu {\psi}_\mu  \right)^2  + \dots 	
\end{eqnarray}
where $\kappa^2 \equiv 8\pi G \equiv \frac{1}{M_{Pl}^2}$ (with $M_{Pl}$ the \emph{reduced} Planck mass) is the gravitational constant, 
$e=\sqrt{-g}$ is the vierbein determinant, 
$R\left[e\right]$ is the scalar curvature and $ D_\mu$ is the gravitational covariant derivative, both in the absence of torsion,
and the dots indicate contributions of the minimal set of auxiliary fields  $\left(A_\mu,\;S,\;P\right)$ required for closure of the 
local supersymmetry algebra, which we do not write explicitly here, as they are of no special interest for our purposes. 

When global supersymmetry is spontaneously broken (\ref{fterm}), 
the gravitino is coupled to the Goldstino field $\lambda$ via the embedding of (\ref{val}) in
the supergravity context, yielding the super-Higgs effect~\cite{deser}.
To see this, we promote the global supersymmetry of (\ref{val}) to a local one, by allowing the parameter $\alpha \left[x\right]$ 
to depend on the space-time coordinates, and couple the action (\ref{val}) to that of $\mathcal{N}=1$ supergravity in such a way 
that the combined action is invariant under the following local supersymmetry transformations:
	\begin{eqnarray}\label{sugratransgolds}
		\delta \lambda &=& f\, \alpha \left[x\right] + \dots ~, \nonumber \\
		\delta e^a _\mu  & = & -i \kappa \overline{\alpha} \left[x\right] \gamma^a \psi_\mu~, \nonumber \\
		\delta \psi_\mu & = & - 2 \kappa^{-1} \partial_\mu \alpha  \left[x\right] + \dots
	\end{eqnarray}
where the $\dots $ in the $\lambda$ transformation denote non-linear $\lambda$-dependent terms (\emph{cf.} (\ref{trnl})). 
The action that changes by a divergence under these transformations is the standard $\mathcal{N}=1$ supergravity action plus
	\begin{equation}\label{va2b}
		L_\lambda \; = \; - f^2 e - \frac{i}{2}\overline{\lambda} \gamma^\mu \partial_\mu \lambda - \frac{i\,f}{\sqrt{2}} \overline{\lambda} \gamma^\nu \psi_\nu + \dots ,
	\end{equation}
which contains the coupling of the Goldstino to the gravitino. 
The Goldstino can be gauged away~\cite{deser} by a suitable redefinition of the gravitino field and the tetrad.
One may impose the gauge condition
	\begin{equation}\label{gravinogauge}
		\psi_\mu \gamma^\mu \; = \; 0 ,
	\end{equation}
but this leaves behind a \emph{negative cosmological constant term}, $-f^2\, e$, so the total Lagrangian after these redefinitions reads:
	\begin{equation}\label{va3}
		\mathcal{L}_{\rm eff} \; = \; -f^2e + (\mathcal{N}=1~{\rm supergravity~ Lagrangian ~ in~Eq.}~(\ref{msugra})). 
	\end{equation}
The presence of four-gravitino interactions in the standard $\mathcal{N}=1$ supergravity Lagrangian
in  the second-order formalism, due to the fermionic contributions to the torsion in the spin connection, 
implies that an induced  gravitino mass term is generated dynamically.

To see this, one may linearize the appropriate four-gravitino mass terms of the $N=1$ supergravity
Lagrangian in the second-order formalism~\cite{ferrara}
by means of an auxiliary scalar field $\rho (x)$~\cite{smith}:
\begin{eqnarray}\label{effactionlinear}
\mathcal{L}{\rm eff} \; = \;
-\frac{1}{2\, \kappa^2} R(e) - \frac{1}{2}\epsilon^{\mu\nu\rho\sigma} \overline{\psi}_\mu \gamma_5 \gamma_\nu D_\rho \psi_\sigma + \rho^2(x) - \sqrt{11} \kappa \rho (x) \left(\overline{\psi}_\mu \Gamma^{\mu\nu} \psi_\nu \right) + \dots
\end{eqnarray}
with $\Gamma^{\mu\nu} \equiv \frac{1}{4} [ \gamma^\mu, \gamma^\nu ]$, where the $\dots $ indicate terms that we are not
interested in, including other four-gravitino interactions with $\gamma_5$ insertions, as well as other
standard $N=1$ interactions and auxiliary supergravity fields.
On account of the gauge-fixing condition (\ref{gravinogauge}) and the anti-commutation properties of the Dirac matrices $\gamma^\mu$, 
we have
\begin{equation}\label{identity} 
 \overline{\psi}_\mu \Gamma^{\mu\nu} \psi_\nu \; = \; -\frac{1}{2}\overline{\psi}_\mu \psi^\mu \, .
\end{equation}
The formation of a condensate
\bea \label{gravinocond}
\langle \rho (x) \rangle \equiv \rho \sim ~\langle \overline{\psi}_\mu \Gamma^{\mu\nu} \psi_\nu \rangle \; \ne \; 0 ~,
 \eea
which should be independent of $x$ because of the translation invariance of the vacuum,
is possible when the effective action  (\ref{effactionlinear}) is minimized along the lines in~\cite{smith}. 
Such a condensate corresponds to a dynamically-generated gravitino mass $M_{3/2}$.
The formation of the condensate may cancel the negative cosmological
constant term~\cite{deser,smith}, since it contributes to the vacuum energy a term of the form
\begin{equation}\label{verho}
\int d^4 x \,e \,\rho^2  > 0~,
\end{equation}
and at tree level one can cancel the negative cosmological constant
of the Volkov-Akulov Lagrangian (\ref{va2b},\ref{va3}), which depends on the supersymmetry breaking scale $f^2$, by setting:
\begin{equation}\label{condition}
\rho^2 =  f^2~.
\end{equation}
In~\cite{smith2}, a one-loop effective potential analysis has demonstrated that such a cancellation occurs for a 
suitable value of the parameter $f$.
Whether the situation persists to higher orders, so that the cancellation of the effective cosmological constant can be achieved exactly,
is not known.

We stress at this stage the important r\^ole of gravitino torsion condensates in providing the appropriate 
cosmological constant terms in the effective action that cancel any bare contribution, leading to
vanishing vacuum energy at the non-trivial minimum of the (one-loop) effective potential. This property is known in 
general relativity as parallelism, and played an important r\^ole in early studies of (spontaneous) 
compactification of higher-dimensional supergravities, such as 11-dimensional supergravity~\cite{duff},
yielding four-dimensional manifolds with zero cosmological constant.

The above considerations have been disputed in~\cite{odintsov}, where it was claimed that, 
in a linearised-gravity approximation, $g_{\mu\nu} = g^{0}_{\mu\nu} + h_{\mu\nu}$, about a de Sitter solution of the 
field equations coming from the one-loop effective action, 
integrating out the metric fluctuations $h_{\mu\nu}$ in the way suggested in \cite{fradkin}
leads to  imaginary parts in the effective action,
indicating an instability of the gravitino condensate. However, these claims were revisited in~\cite{alexandre},
where it was found that there are solutions corresponding to non-trivial minima of the effective potential where 
such imaginary parts are not present. In fact, the corresponding  one-loop effective potential computed with the 
quantum-gravitational fluctuations has exactly the double-well shape of the  potential of~\cite{smith2} in flat space-time, 
supporting the claim of~\cite{smith2} for the possibility of dynamical breaking of local supergravity. 

In the present work we revisit such a scenario for ${\mathcal N}=1$ dynamical
supergravity breaking with a view to inflation, with the r\^ole of the inflaton being played by the gravitino condensate field
that is responsible for providing a mass for the gravitino field via its v.e.v..
We stress that, in the dynamical supergravity breaking scenarios of standard supergravity discussed in~\cite{smith,smith2}, 
the gravitino mass is of the order of Planck mass. However, for the purpose of our analysis we require
a gravitino that is light compared to the Planck mass scale. 
In order to obtain such a light dynamical gravitino masse one should consider extended versions of ${\mathcal N=1}$ supergravity
coupled conformally to additional fields.  

One such extension was discussed 
in~\cite{emdyno}, where we extended ${\mathcal N}=1$ supergravity to include a Barbero-Immirzi field, which was identified with a 
complex chiral superfield coupled conformally to the supergravity action. The scalar component of this superfield is identified with a 
complex scalar field, whose real part is the dilaton responsible for breaking of (super)conformal symmetry, and whose imaginary part
is an axion associated with a field extension of the Barbero-Immirzi parameter~\cite{gates}.
The fermionic Majorana spin-1/2 component of this chiral superfield (dilatino)
can be identified with the Goldstino field, whose infrared behaviour is described  by the Volkov-Akulov Lagrangian (\ref{val}),
with a coupling to the  ${\mathcal N}=1$ supergravity action given by (\ref{va2b}). 
As in standard supergravity, the Goldstino field is eaten by the appropriate component of the gravitino, 
as in (\ref{gravinogauge}), and its only remnant is a negative cosmological constant $-f^2$ in the effective action. 
We assume that the dilaton is stabilized by minimisation of an appropriate potential whose details are not specified. 
Since broken global supersymmetry implies a non-zero (positive) non-trivial minimum for the dilaton potential, 
such terms are absorbed in $f^2$.

The important difference from the standard supergravity action induced by such a stabilized dilaton is  that the 
coupling constants of the four-gravitino interactions, in the Einstein frame where the scalar curvature term and the 
kinetic term of the gravitino are canonically normalized, 
is no longer the gravitational coupling $\kappa^2 $, but a scaled one, involving multiplicative factors of the
dilaton v.e.v.. In particular, the relevant supergravity terms in the 
Einstein frame (denoted by a superscript E) now read~\cite{ferrara2,emdyno}
 \begin{eqnarray}\label{confsugra2}
\mathcal{L}^E  (e^E)^{-1} & = &  -\frac{1}{2\kappa^2} R^E(e^E)  - \frac{1}{2} \epsilon^{\mu\nu\rho\sigma} \overline{\psi '}_\mu  \gamma_5 \gamma_\nu D^E_\rho {\psi'}_\sigma -  e^{2\varphi}\,V^E
 - \nonumber \\ &~& \frac{11 \kappa^2}{16} e^{-2\varphi} \left[ (\overline{\psi'}_\mu  {\psi'}^\mu )^2  - (\overline{\psi'}_\mu  \gamma_5 {\psi'}^\mu )^2 \right]
+ \frac{33}{64} \kappa^2 e^{-2\varphi} \, \left(\overline{\psi'}^\rho \gamma_5  \gamma_\mu {\psi'}_\rho  \right)^2  + \dots   \nonumber \\
& = &
 -\frac{1}{2\kappa^2} R^E(e^E)  - \frac{1}{2} \epsilon^{\mu\nu\rho\sigma} \overline{\psi '}_\mu  \gamma_5 \gamma_\nu D^E_\rho {\psi'}_\sigma -  e^{2\varphi}\,V^E
 +  \nonumber \\ &~&   \rho^2(x)  +  \frac{\sqrt{11}}{2}\kappa \rho (x) e^{-\varphi}\, \left(\overline{\psi'}_\mu {\psi'}^\mu \right)  + \pi^2(x)  +  \frac{\sqrt{11}}{2}e^{-\varphi}\, \kappa \, i \pi (x)  \left(\overline{\psi'}_\mu \gamma_5 {\psi'}^\mu \right) + \nonumber \\ &~& \frac{\sqrt{33}}{2} \kappa \, e^{-\varphi}\, i \lambda^\nu \left(\overline{\psi'}^\rho \gamma_5  \gamma_\nu {\psi'}_\rho \right) + \dots ~,
 \end{eqnarray}
where $\varphi$ denotes the (constant in space-time) dilaton v.e.v. that breaks conformal symmetry and global supersymmetry, 
$\psi'_\mu$ denotes the canonically-normalized gravitino with a standard kinetic term as in $\mathcal{N}=1$ supergravity,
and the $\dots$ denote structures, including auxiliary fields, that are not of interest here. In writing
(\ref{confsugra2}) we have expanded the four-gravitino terms into detailed structures to exhibit explicitly the terms that
generate masses, and we have linearized the four-gravitino terms. The condensate of interest to us is the v.e.v. of the 
linearizing field $\rho (x)$, $\langle \rho (0) \rangle$, which is independent of space-time coordinates, because 
of translation invariance.  As already mentioned, the 
minimum of the dilaton potential contributes to the cosmological constant, and the terms $e^{2\varphi} V^E $ above
are identified with the (negative) cosmological constant $-f^2$ associated with the scale of global supersymmetry breaking
that appears when the goldtsino field is coupled to the Lagrangian (\ref{confsugra2}), as in (\ref{va3}). 

Due to the presence of the dilaton v.e.v. $\varphi$, the Lagrangian (\ref{confsugra2}) has an
extra phenomenological parameter as compared to the standard supergravity case, namely the interaction 
constant in front of the four-gravitino terms:
\begin{equation}\label{kaptilde}
{\tilde \kappa} \; = \; e^{-\varphi} \, \kappa ~.
\end{equation}
In the absence of a non-trivial dilaton v.e.v., $\varphi = 0$, the supergravity Lagrangian reduces to the 
standard ${\mathcal N}=1$ Lagrangian and the confornal coupling ${\tilde \kappa} $ becomes identical to the gravitational constant $\kappa$. 
We considered in~\cite{emdyno} the case where ${\tilde \kappa} \gg \kappa $,
in which case the quantum-gravitational fluctuations of the metric field 
could be safely ignored, and performed a consistent analysis of dynamical gravitino mass generation in Minkowski space-time.
As we show in this work, such a situation also favours inflationary phenomenology. 

In the next Section~\ref{sec:one loop},  we construct the one-loop effective action of  conformal supergravity models 
and study the formation of gravitino condensates with mass scales much lighter than the Planck mass. 
In Section~\ref{sec:infl}  we argue for inflation via the effective potential of the condensate field around its origin, 
and perform a phenomenological analysis finding consistency with the Planck~\cite{Planck} CMB data. 

\section{Analysis of the One-Loop Effective Potential \label{sec:one loop} \label{sec:oneloop}}

After integrating out the gravitino field in a flat Minkowski space-time, 
to a one-loop effective potential. This can be 
obtained by following the analysis of~\cite{smith2}, but replacing the gravitational coupling in 
that work by the new coupling $\tilde{\kappa}$ (\ref{kaptilde}).  In a Minkowski space-time the effective potential is divergent in the ultraviolet, 
so to regularize this divergence one needs to impose a UV momentum cutoff $\Lambda$, which is a phenomenological parameter 
in our inflationary scenario, as we discuss below. This sets the scale of inflation and also of the associated dynamical 
mass of the gravitino that is obtained at the end of the inflationary period, as we also see below. 
Keeping just the lowest-order terms in the derivative expansion,
the one-loop effective action takes the generic form
\begin{equation}\label{effaction2}
\Gamma = \int d^4 x \sqrt{-g} \Big (Z[\rho] \, \partial_\mu \, \rho \, \partial^\mu \, \rho - {\mathcal V}_{\rm eff} [\rho ] \Big) + \dots \, ,
\end{equation}
where where the $\dots$ denote higher-order derivatives, $Z[\rho]$ is the wave function renormalization, and ${\mathcal V}_{\rm eff}$ is the effective potential for the $\rho $ field, defined as ${\mathcal V}_{\rm eff} = - \Gamma [{\rm nonderivative~terms~in}\, \rho] $. 

\begin{figure}[ht]
\centering
\vspace{-0.4cm}
\includegraphics[width=12cm]{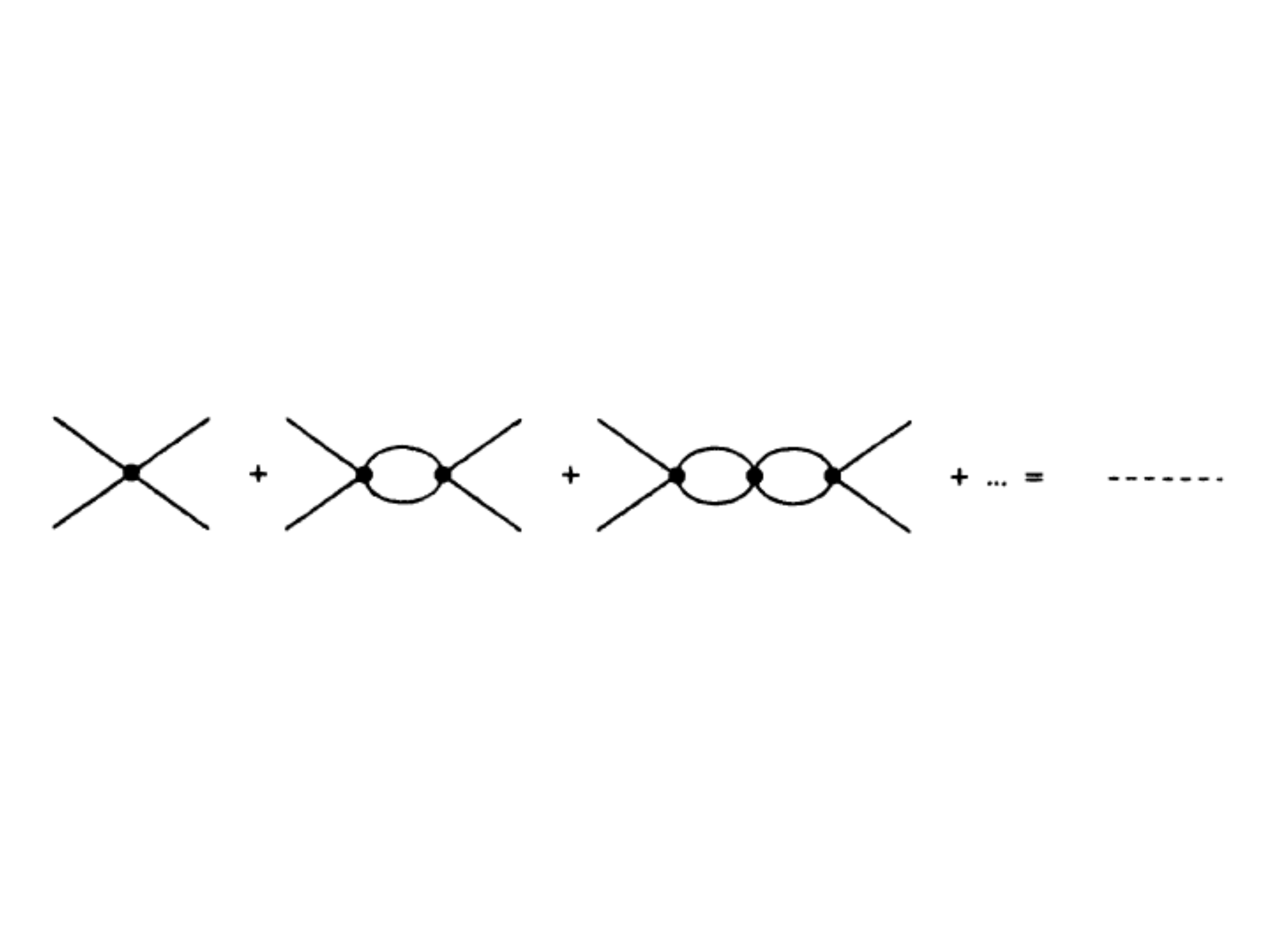} 
\vspace{-3cm}
\caption{\it Generic diagrammatic structure of the scalar-channel bubble four-gravitino Feynman graphs (continuous lines) 
contributing to the condensate-field $\rho$ propagator (dashed line) in ${\mathcal N}=1$ supergravity.}%
\label{fig:channel}%
\end{figure}

The  (one-loop) wave-function renormalization $Z[\rho]$, which promotes the initially auxiliary condensate field $\rho$ into 
a dynamical one, can be obtained by similar diagrammatic methods as used in the analysis
of composite Higgs models~\cite{bardeen}, thanks to a close analogy of the dynamical local supersymmetry breaking 
with the dynamical breaking of gauge symmetries in those models. The gravitino condensate field plays the same r\^ole 
as the top-quark (fermion) condensate field in~\cite{bardeen}. Specifically, one may use split the condensate field 
into its classical vacuum expectation value (v.e.v.) and quantum $\tilde \rho$ parts:
$\rho = \langle \rho \rangle + {\tilde \rho}(x)$, where $ \langle \rho \rangle $ yields a bare gravitino mass. 
Using the massive gravitino propagator in flat space~\cite{VanNieuwenhuizen:1981ae}, one may 
write a Schwinger-Dyson-like gap equation for the gravitino field, and construct the propagator for
the condensate field $\rho$ by summing the appropriate scalar channel bubble diagrams generated using
the four-fermion gravitino terms in the supergravity action, as depicted in Fig.~\ref{fig:channel}.
 
The details are not relevant for our purposes here, the important point is the UV-divergent structure of the 
wave-function renormalization factor $Z[\rho]$. Using the Fourier transform of a massive gravitino 
propagator~\cite{VanNieuwenhuizen:1981ae}:
\begin{equation}\label{massgravino}
{\tilde P}_{ab} = - \Big[ \, \big(\delta_{ab} + p_a \, p_b /M_{3/2}^2 \big) \, \big(i \, \gamma^\mu \, p_\mu  - M_{3/2}\big)
+ \frac{1}{3} \, \big(\gamma_a - i \, p_a/M_{3/2} \big)\, \big(i \, \gamma^\mu \, p_\mu  + 
M_{3/2}\big) \, \big(\gamma_b - i \, p_b/M_{3/2} \big) \, \Big] \, /\, \Big(p^2 + M_{3/2}^2\Big ) ~,
 \end{equation}
where $M_{3/2} \propto \langle \rho \rangle $ is the gravitino mass and, following a similar analysis as in~\cite{bardeen}, 
one may derive the scalar-condensate propagator, 
from which one can deduce the UV structure of the wave-function renormalization in (\ref{effaction2}). The
bubble graphs of Fig.~\ref{fig:channel}  may be resummed to yield the propagator $\Gamma_s(p)$ of the scalar condensate, 
which has the generic structure
\begin{equation}
\Gamma_s(p) \; \sim \; \frac{i}{p^2 + M_\rho^2} \times \Big[ {\mathcal O}(\Lambda^4) + {\mathcal O}(\Lambda^2)  +  \mathcal{O}\big({\rm ln}\big(\frac{\Lambda}{\mu}\big)^2\Big) \Big]^{-1} \, ,
\label{prop}
\end{equation}
where we have separated the various terms according to their UV-divergent structure, namely 
quartic, quadratic and logarithmic UV divergences, and $M_\rho$ denotes the condensate mass.
In the dynamical Higgs scenario of~\cite{bardeen}, the resummation of the bubble graphs of Fig.~\ref{fig:channel}
would lead to the prediction that the condensate mass is twice the top-quark mass. Renormalization-group studies then show
that the condensate mass is actually slightly higher than the renormalized top-quark mass, and is cutoff dependent in general. 
In our case the situation is more complicated, since there are various types of four-fermion interactions that should
be taken into account, all with similar strength $\sim {\tilde \kappa}^2$.  We do not perform 
their full resummation here, being content with the estimate of the condensate mass
made in the next Section, based on the minimization of the one-loop effective potential. 
Here we sketch the analysis, concentrating on the derivation of the logarithmic scaling 
of the wave-function-renormalization with the cutoff scale. 
Using the appropriate gap equation for the dynamically-generated gravitino mass~\cite{smith},  
namely
\begin{equation}
\frac{3}{44 \, {\tilde \kappa}^2} \; = \; \int^\Lambda d^4 k \Big( 4 + k^2/M_{3/2}^2 \Big) \frac{1}{k^2 + M_{3/2}^2} 
\; = \; \pi^2 \Big[ \Lambda^4/(2 M_{3/2}^2) + 3 \Lambda^2 - 3M_{3/2}^2 \, {\rm ln}\big(\Lambda^2/M_{3/2}^2 + 1 \big)\,\Big] \, ,
\label{gap}
\end{equation}
arising from the requirement of the vanishing of the tadpole of the quantum field ${\tilde \rho}(0)$~\cite{smith}, one observes that 
the power UV divergences inside the brackets in the above expression for $\Gamma_s(p)$ cancel out, leaving only the 
logarithmic terms to contribute to the wave-function renormalization $Z[\rho]$.   
Using a transmutation scale $\mu$, the latter can be written generically as: 
\begin{equation}
Z[\rho ] \sim {\rm const.} \times {\rm ln}\Big(\frac{\Lambda^2}{\mu^2}\Big) \, ,
\end{equation}
where the proportionality constant denotes numerical coefficients whose precise value is not relevant for our purposes 
(we note, though, that such coefficients contain factors $1/(4\pi)^2 \sim {\mathcal O}\big(1/100 \big)$). 
The canonically-normalized scalar condensate field $\phi$ that plays the r\^ole of the inflation is then:
\begin{equation}\label{canonical}
\phi (x) \equiv \sqrt{2 \, \times \, {\rm const.}  \, \times {\rm ln}\, \Big(\frac{\Lambda^2}{\mu^2}\Big)}\; {\tilde \kappa}\, \rho (x)~.
\end{equation}
Using the standard propagator of the (masssless) gravitino 
field~\cite{VanNieuwenhuizen:1981ae} in Minkowski space-time,
the one-loop effective potential for the field $\rho(x)$ at one-loop order is then found to be~\cite{smith2}:
\begin{eqnarray}\label{effpotsugra}
e^{-1} \, \mathcal{V}_{\rm eff} \; & = & \;  \int d^4 x \, \sqrt{-g} \, \Big( {\rm Lim}_{V \to \infty } \Big[ \frac{1}{2\, V} \, \sum_{n=1}^{\infty} \, \frac{\Big(\sqrt{11} {\tilde \kappa}\Big)^{2n}}{2n} \, {\rm Tr} \big[P_{ab} \rho \big]^{2n} \Big]  + f^2 - \rho^2 \Big) \, 
\nonumber \\
& = & \frac{4}{(2\pi)^4} \int^\Lambda  d^4 p \, 
{\rm ln}\left(1 + 11 \frac{{\tilde \kappa}^2 \rho^2}{p^2} \right) + f^2 - \rho^2  = 
(\frac{1}{4\pi^3})\{ \frac{121}{2} {\tilde \kappa}^{\prime 4} \rho^4 \left[
{\rm ln}(11 {\tilde \kappa}^{\prime 2} \rho^2 /\Lambda^{\prime 2}) - \frac{1}{2} \right]
+ 11 {\tilde \kappa}^{\prime 2} \rho^2 \Lambda^{\prime 2} \} + f^2 - \rho^2 ~, \nonumber \\
& = & ({\tilde \kappa}^{\prime})^{-4} \Big[ (\frac{1}{4\pi^3})\{ \frac{121}{2}  \sigma ^4 \left[
{\rm ln}\Big(\frac{11 \, \sigma ^2}{{\tilde \kappa}^{\prime 2}\Lambda^{\prime 2}}\Big) - \frac{1}{2} \right]
+ \sigma ^2 \big(\frac{11}{4\pi^3} {\tilde \kappa}^{\prime 2} \Lambda^{\prime 2}  -1 \Big) + f^2 {\tilde \kappa }^{\prime 4} \Big] ~,
\end{eqnarray}
where ${\tilde \kappa}^\prime \equiv {\tilde \kappa}\, (\pi)^{1/4}, \Lambda^\prime \equiv \Lambda (\pi)^{1/4}$.
This rescaling has been performed in order for the one-loop corrections of the effective potential to have the same form as in~\cite{smith2}, 
where such $\pi$ factors were lacking.
For brevity, in the remainder of this article we use the unprimed notation for both ${\tilde \kappa}$ and $\Lambda$, 
with the understanding that now ${\tilde \kappa}$ is given by
\begin{equation}\label{kaptildenew}
{\tilde \kappa} \; = \; e^{-\varphi} \, (\pi)^{1/4} \, \kappa = e^{-(\varphi - \frac{1}{4}\,{\rm ln}\pi)} \, \kappa  ~,
~\end{equation}
instead of (\ref{kaptilde}). In (\ref{effpotsugra}), $V \to \infty$ is a space-time volume, which cancels the $\int d^4 x $ factor for constant fields $\rho$ (as appropriate for the evaluation of the effective potential) and $P_{ab}$ is the (massless) 
gravitino propagator, which, after using the gauge condition (\ref{gravinogauge}), becomes:
\begin{equation}\label{gravprop}
P_{ab} \; = \; - \frac{1}{2} \, \frac{\gamma_b \gamma^\mu \, \partial_\mu \, \gamma_a }{\Box} \, ,
\end{equation} 
where $\Box \equiv \partial^\mu \, \partial_\mu$ is the d'Alembertian. Its Fourier transform reads
\begin{equation}\label{prop}
 {\tilde P}_{ab} \; = \;  \frac{1}{2} i\, \frac{\gamma_b \gamma^\mu \, p_\mu \, \gamma_a }{p^2} \, ,
\end{equation} 
which has been used to obtain the final expression for the effective potential given in the last line of (\ref{effpotsugra}), which we shall use in our discussion of inflation in the next section. 
The field $\sigma $ is a dimensionless rescaled condensate field, $\rho^2 \to {\tilde \kappa }^{-4} \, \sigma ^2 $, or, in terms of $\phi$:
\begin{equation}\label{sphi}
\sigma = \Big(\frac{{\tilde \kappa}} {\kappa}\Big)  \, \frac{1}{\sqrt{2\, {\rm const} \times \, {\rm ln}\Big(\frac{\Lambda^2}{\mu^2}\Big)}} \, \kappa \, \phi. 
\end{equation}

The expression (\ref{effpotsugra}) contains the following undetermined parameters that are to be constrained by observation:

~\\

$\bullet$  The cutoff $\Lambda$, which appears in the one-loop effective potential through the dimensionless 
combination ${\tilde \kappa^2} \Lambda^2 $ and in the regularized logarithmic divergence of the wave-function 
renormalization of the kinetic term of the one-loop effective action through the dimensionless combination 
${\rm ln}~\Big(\Lambda^2/\mu^2\Big)$, where $\mu$ is a transmutation scale;

~\\

$\bullet$ The overall scale of the effective potential ${\tilde \kappa}^{-4}$, which may be adjusted by varying 
the v.e.v. of the dilaton field;

~\\

$\bullet$ The scale of the cosmological constant $f^2$, which is related to the scale of global supersymmetry breaking.

~\\

\begin{figure}[ht]
\centering
\includegraphics[width=9cm]{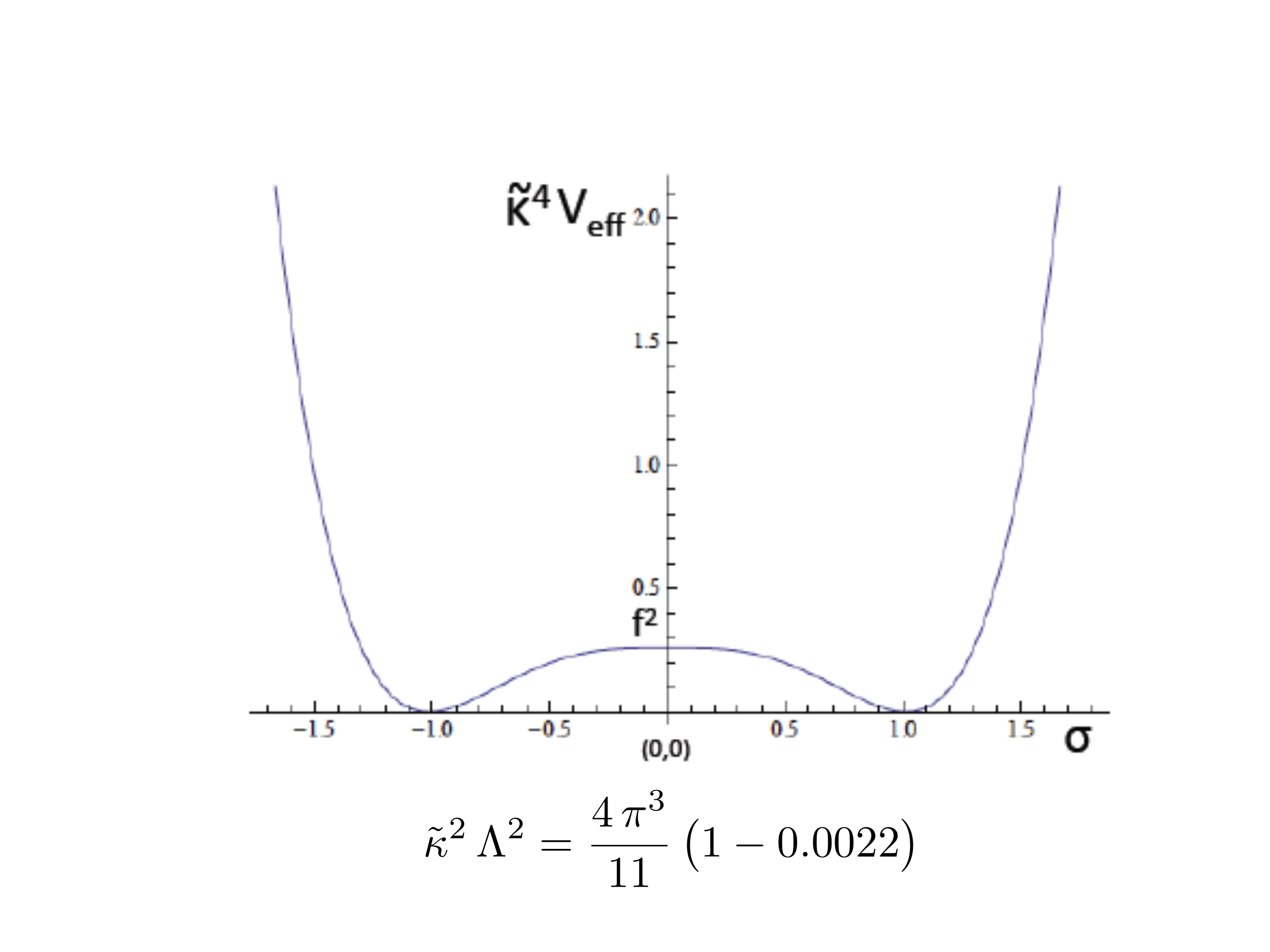} \hfill \includegraphics[width=7.5cm]{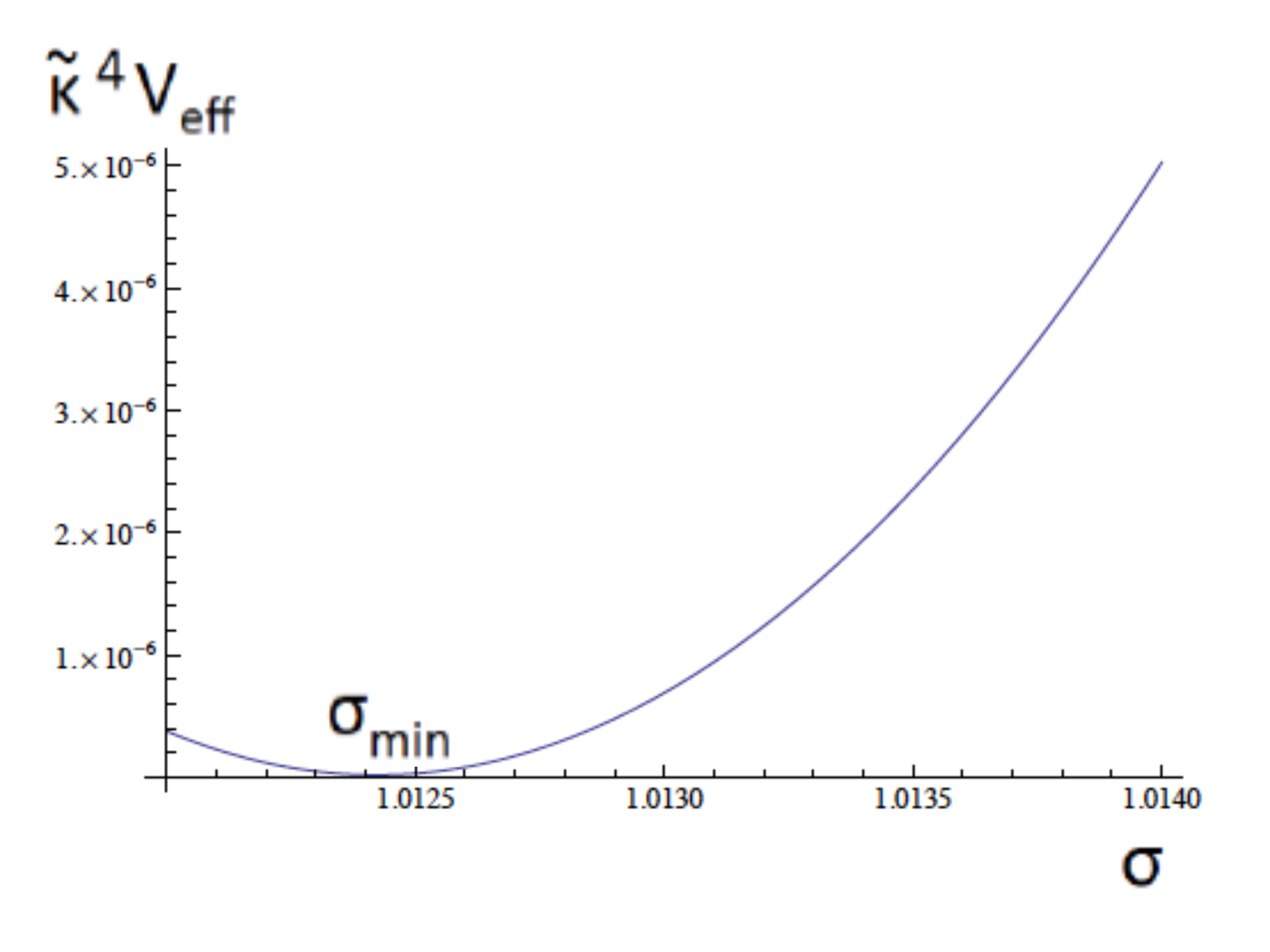}
\caption{\it \emph{Left Panel:} The one-loop effective potential (\ref{effpotsugra})  for the field $\rho(x)$ in the 
one-loop effective action (\ref{confsugra2}) of the ${\mathcal N}=1$ conformal supergravity, which is suitable for 
dynamical breaking of local supersymmetry and the generation of a gravitino mass. \emph{Right Panel:}
For the indicated values of the relevant parameters the effective potential (which is symmetric about the origin)
vanishes at its non-trivial minima: $\sigma = \sigma_{\rm min}$, corresponding to a vanishing effective
cosmological constant at one-loop order.}%
\label{fig:effpot}%
\end{figure}

\begin{figure}[ht]
\centering
\includegraphics[width=5.5cm]{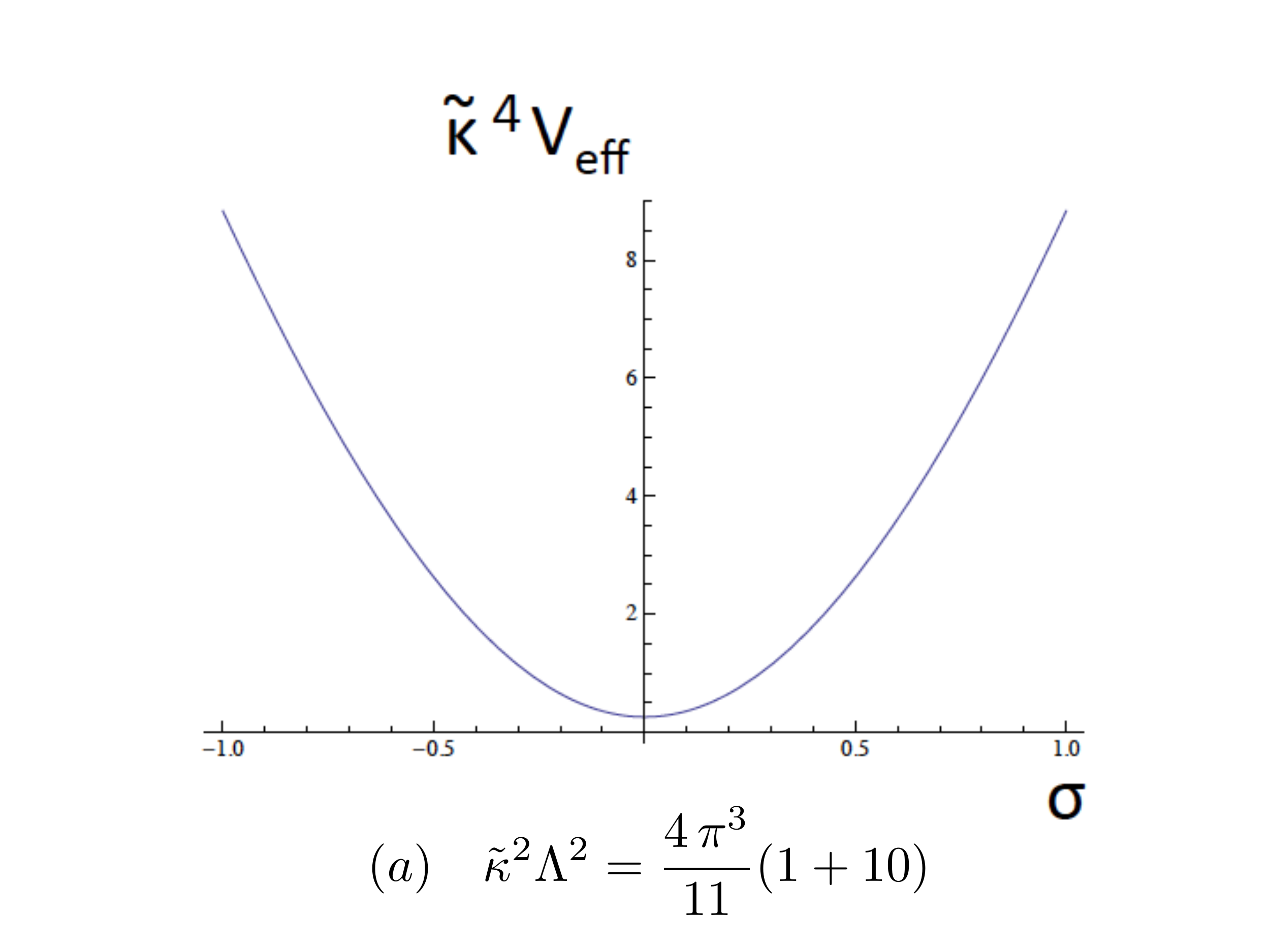} \hfill \includegraphics[width=5.5cm]{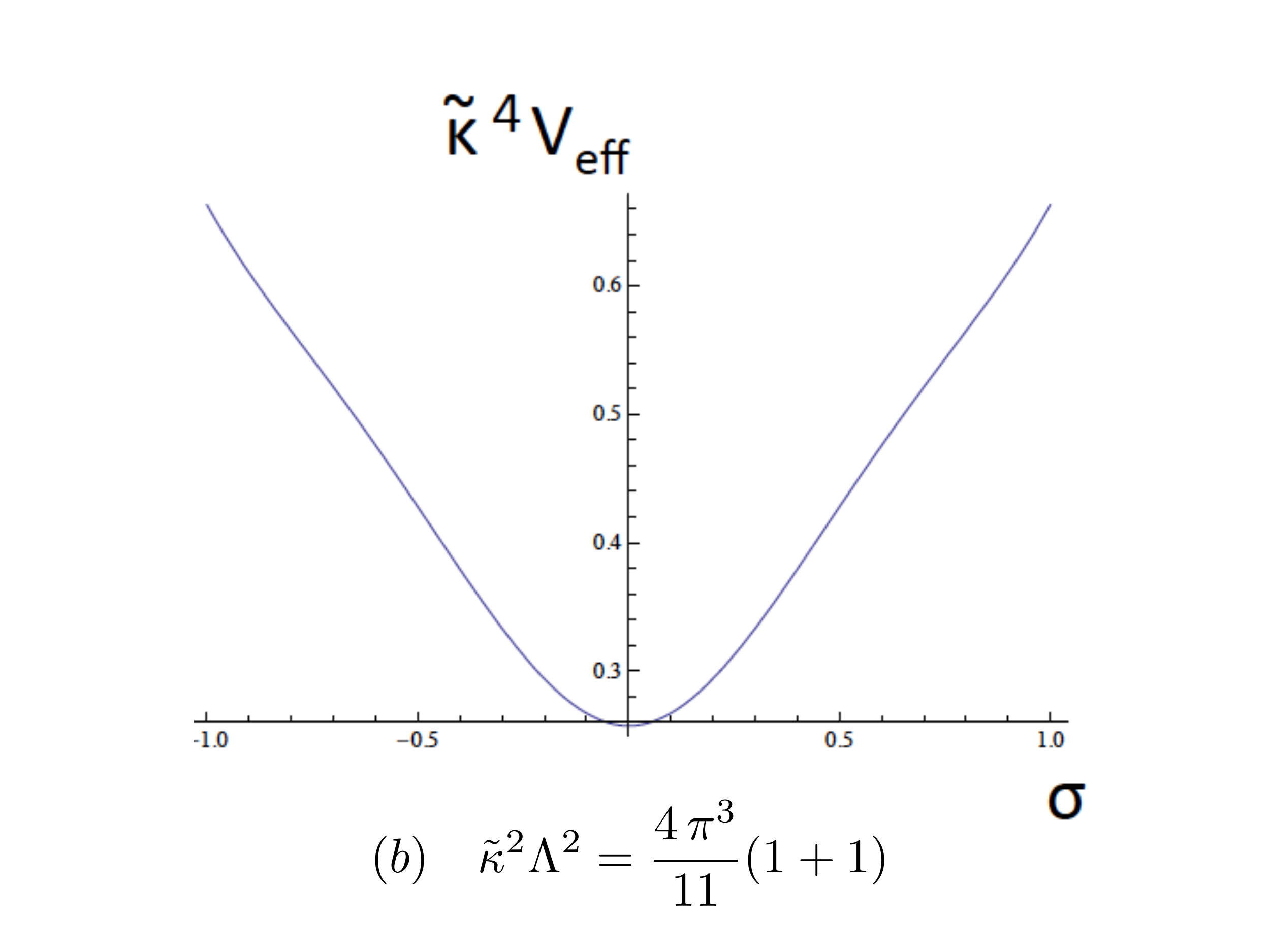} \hfill \includegraphics[width=5.5cm]{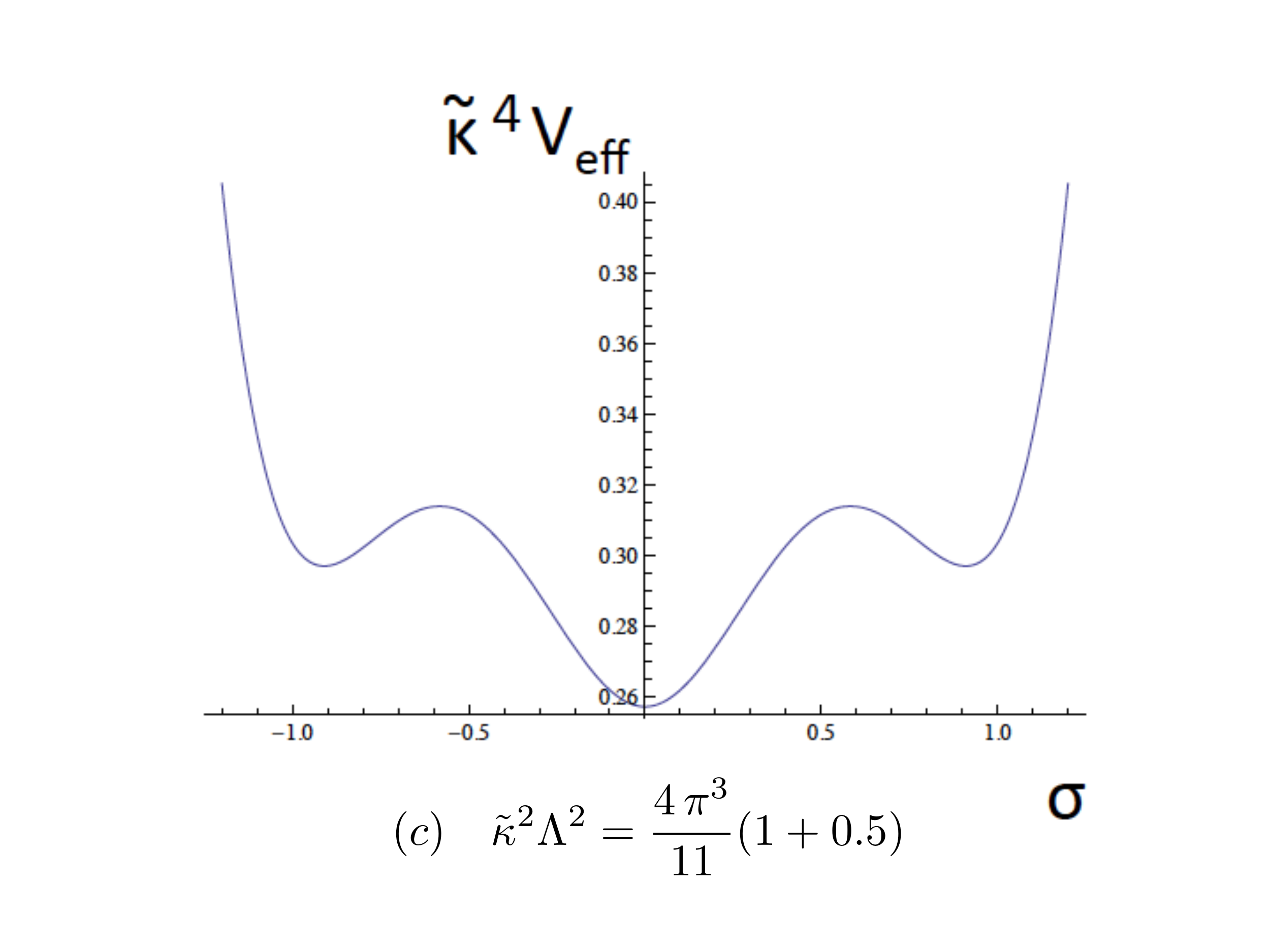} 
\caption{\it The shape of the one-loop effective potential (\ref{effpotsugra}) for various values of the product 
${\tilde \kappa}^2 \, \Lambda^2 $ as the cutoff runs from the UV to the IR, before symmetry breaking. 
Decreasing ${\tilde \kappa}^2 \, \Lambda^2 $ further yields the inflationary potential shown in the
left panel of Fig.~\ref{fig:effpot}. }%
\label{fig:various}%
\end{figure}

The appearance of the super-Higgs effect guarantees the existence of the third parameter, 
which is important for ensuring 
that the effective potential is non-negative for certain values of $\Lambda$ and $f^2$, as we discuss below,
and has a minimum at zero cosmological constant as seen in Fig.~\ref{fig:effpot}, allowing 
us to interpret terms of the form $<\rho> {\overline \psi}'_\mu {\psi'}^\mu $ as corresponding to gravitino mass terms~\cite{emdyno,smith2}
in Minkowski space-time.
In general, the shape of the potential depends on the value of the cutoff $\Lambda$ relative to the coupling ${\tilde \kappa}$. 
Specifically, as the combination ${\tilde \kappa}^2 \, \Lambda^2 $ becomes smaller, the potential develops 
the non-trivial minima and dynamical breaking of supergravity and inflation can occur. 
In Fig.~\ref{fig:various} we display various shapes of the potential before local supersymmetry breaking for indicative 
values of  ${\tilde \kappa}^2 \Lambda^2$ from large to small.

The reader might worry that, by ignoring the quantum gravitational fluctuations, 
we may arrive at an inconsistent result for the behaviour of the effective potential, but
this is not the case. As demonstrated in~\cite{alexandre}, a rigorous incorporation of such metric 
fluctuations into the effective action, following~\cite{fradkin}, 
still produces effective potentials of the form of Fig.~\ref{fig:effpot}. In such a case, instead of formulating the effective 
theory in flat Minkowski space-times, we 
consider Euclidean de Sitter backgrounds, with a one-loop renormalized cosmological constant ${\tilde \Lambda} > 0$.
The quantity $\sqrt{1/{\tilde \Lambda}}$  now plays the r\^ole of the UV regulator, and in this case one may consider
minimization of the effective action in the Minkowski space-time limit where ${\tilde \Lambda} \to 0$, 
while $f^2$ can be adjusted appropriately so that the value of the one-loop effective potential 
(including the effects of quantum gravity fluctuations) at the non-trivial minimum of the condensate 
$\rho$ vanishes as in the case of~\cite{smith2}. A detailed analysis performed in~\cite{alexandre} 
confirms the existence of such solutions, without the presence of imaginary parts in the one-loop effective action, 
contrary to the claims in~\cite{odintsov}.

\section{An Inflationary Scenario based on Gravitino Condensation \label{sec:infl}} 

An important feature of the effective potential (\ref{effpotsugra}), or those derived in~\cite{alexandre} with 
the inclusion of metric fluctuations about a de Sitter background, 
is its flatness around the local maximum, where the potential is approximately constant and ${\cal O}(f^2)$. 
We now discuss scenarios exploiting this flatness of the effective potential
to drive an inflationary cosmological phase, which ends before the condensate field 
$\rho$  rolls down to the non-trivial minimum at which the value of the effective potential vanishes. 
We see now the importance of a non-trivial super-Higgs effect, 
because it is the v.e.v. of the F-term of the global supersymmetry breaking chiral field, $f^2$, 
that determines the size of the Hubble parameter during the inflationary era.
However, in our conformal supergravity scenario,the overall dilaton factor appearing in 
${\tilde \kappa} $ (\ref{kaptildenew}) also contributes to setting the scale in terms of the actual Planck mass, 
which is important for the  phenomenological exploring of CMB constraints on this inflationary potential, as we discuss below. 

We are interested in an inflationary phase at small condensate fields $\phi (x)$ or, equivalently, 
$\sigma (x)$) around the local maximum of the potential, 
where the latter assumes an approximately constant value of order $f^2 $. 
The slow-roll conditions for the inflaton (gravitino condensate) field  are valid in a region on either side of the origin, 
as we now show. 

The relevant slow-roll inflationary parameters are defined in terms of the potential $V$ by~\cite{encyclopaedia}:
\begin{equation}\label{slowroll}
\epsilon \; = \; \frac{1}{2} M_{Pl}^2 \left( \frac{V'}{V} \right)^2~, \;
\eta \; = \; M_{Pl}^2 \left( \frac{V''}{V} \right)  \, , \quad  \xi \;  =  \; M_{Pl}^4 \left( \frac{V'V'''}{V^2} \right)  \,  ~,
\end{equation}
where the primes denote differentiation with respect to the canonically-normalized inflaton field $\phi$ (\ref{canonical}),
which yield the following observables:
\begin{eqnarray}\label{ns}
{\rm Tensor/Scalar~ratio}: \; r & = & 16 \epsilon~,  \nonumber \\  
{\rm Scalar~Spectral~index}: \;  n_s & = & 1 - 6 \epsilon + 2 \eta ~,
\end{eqnarray} 
For completeness, we also consider the 
running of the spectral index, $\alpha_s \equiv d n_s / d {\rm ln} k$, which affects the scalar
power spectrum~\cite{encyclopaedia} 
\begin{equation}\label{powerspectrum}
P(k) \; = \; A \, {\rm exp}\Big[ (n_s -1){\rm ln}(k/k_0) + \frac{1}{2} \alpha_s \, {\rm ln}^2 (k/k_0)\Big] ~, \quad \end{equation}
where $k_0$ is a pivot point, typically taken to have the value $k_0=0.002$.
In terms of the slow-roll parameters, $\alpha_s$  is given by:
\begin{equation}\label{alphas}
\alpha_s \; = \;  \frac{1}{8\, \pi^2} \Big[ - \frac{\xi}{4} + 2 \, \eta\,  \epsilon - 3\, \epsilon^2 \Big] ~,
\end{equation}
This is in principle an important ambiguity in fits to the CMB data:
for example, the general inflationary fit to the Planck data yields $\alpha_s = - 0.0134 \pm 0.0090$~\cite{Planck},
which is compatible with zero at the 1.5-$\sigma$ level.
However, $\alpha_s$ is expected to be very small in generic slow-roll models. Below we verify that this is indeed the case 
in our gravitino-condensate model, so that  its predictions 
can be confronted with constraints obtained from the data assuming that $\alpha_s \simeq 0$.
The magnitude of the primordial density perturbations imposes the constraint
\begin{equation}
\left(\frac{V}{\epsilon}\right)^{\frac{1}{4}} \; = \;  0.0275 \times M_{Pl} \, 
\label{vconstr}
\end{equation}
on the value of the inflationary potential~\cite{encyclopaedia}.
Finally, the number of e-foldings for the duration of the slow roll, i.e., while the inflaton field
has values in the interval $0 < \phi_i \, < \, \phi \, < \, \phi_e $  is given by:
\begin{equation}\label{efoldings}
N = - \frac{1}{M^2_{Pl}}\, \int_{\phi_i}^{\phi_e} \frac{V}{V^\prime} d\phi ~.
\end{equation}
Motivated by the need to consider asymptotically-flat Minkowski space-time at the end of 
the inflationary epoch, where the gravitino acquires a mass dynamically,
we make the following important observation. If we set in (\ref{sphi}) 
\begin{equation}\label{kappalog}
\frac{{\tilde \kappa}}{\kappa} \; \sim \; \sqrt{2\, {\rm const} \times \,   {\rm ln}\Big(\frac{\Lambda^2}{\mu^2}\Big) } \, \gg \, 1 ~,
\end{equation}
then the dimensionless field $\sigma$ becomes identical to the dimensionless
field $\phi/M_{Pl}$ that plays the r\^ole of the inflation. 
The condition (\ref{kappalog}) is consistent  with ignoring the quantum fluctuations of the graviton field. 

\begin{figure}[ht]
\centering
\includegraphics[width=8cm]{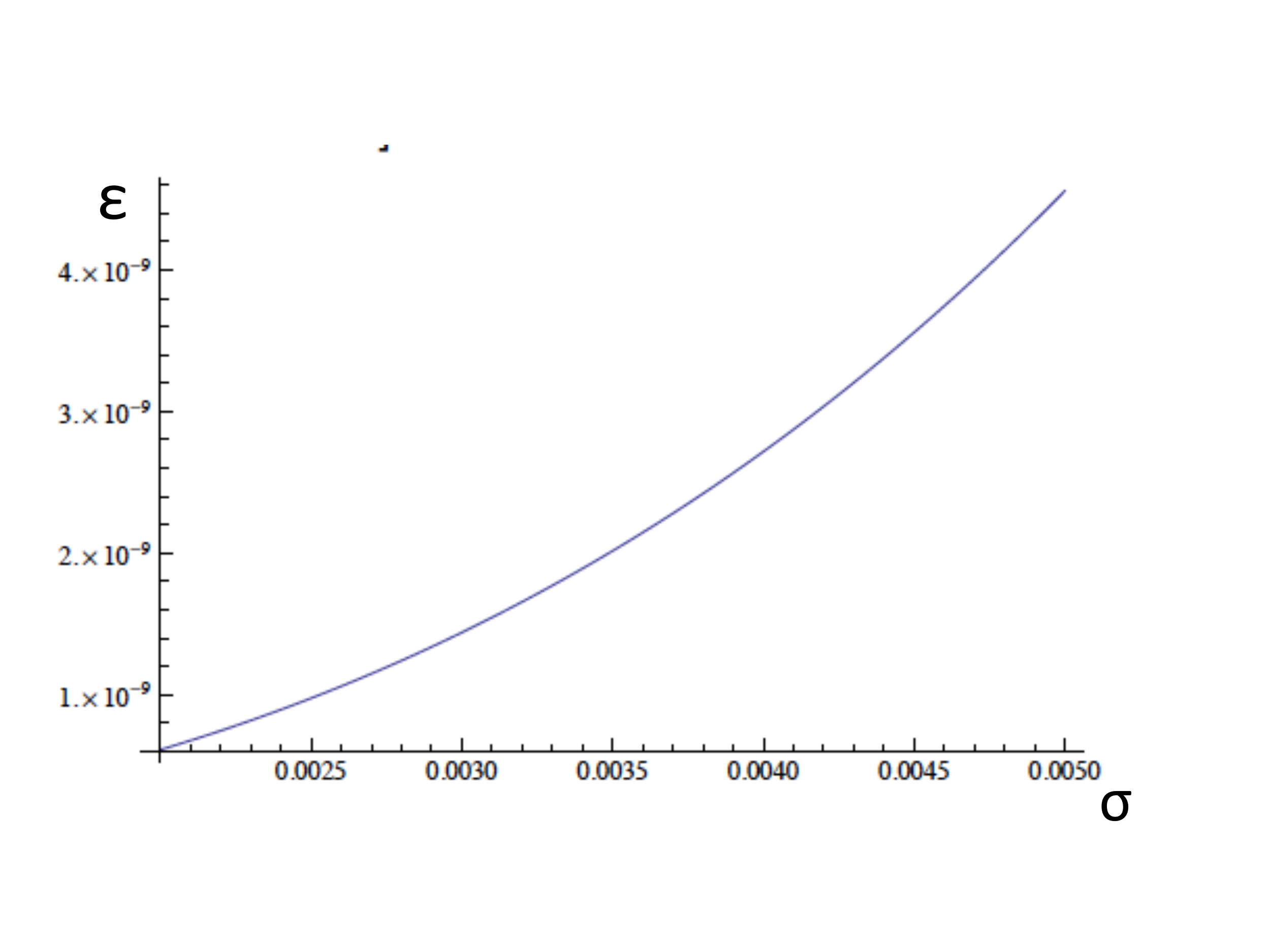} \hfill \includegraphics[width=8cm]{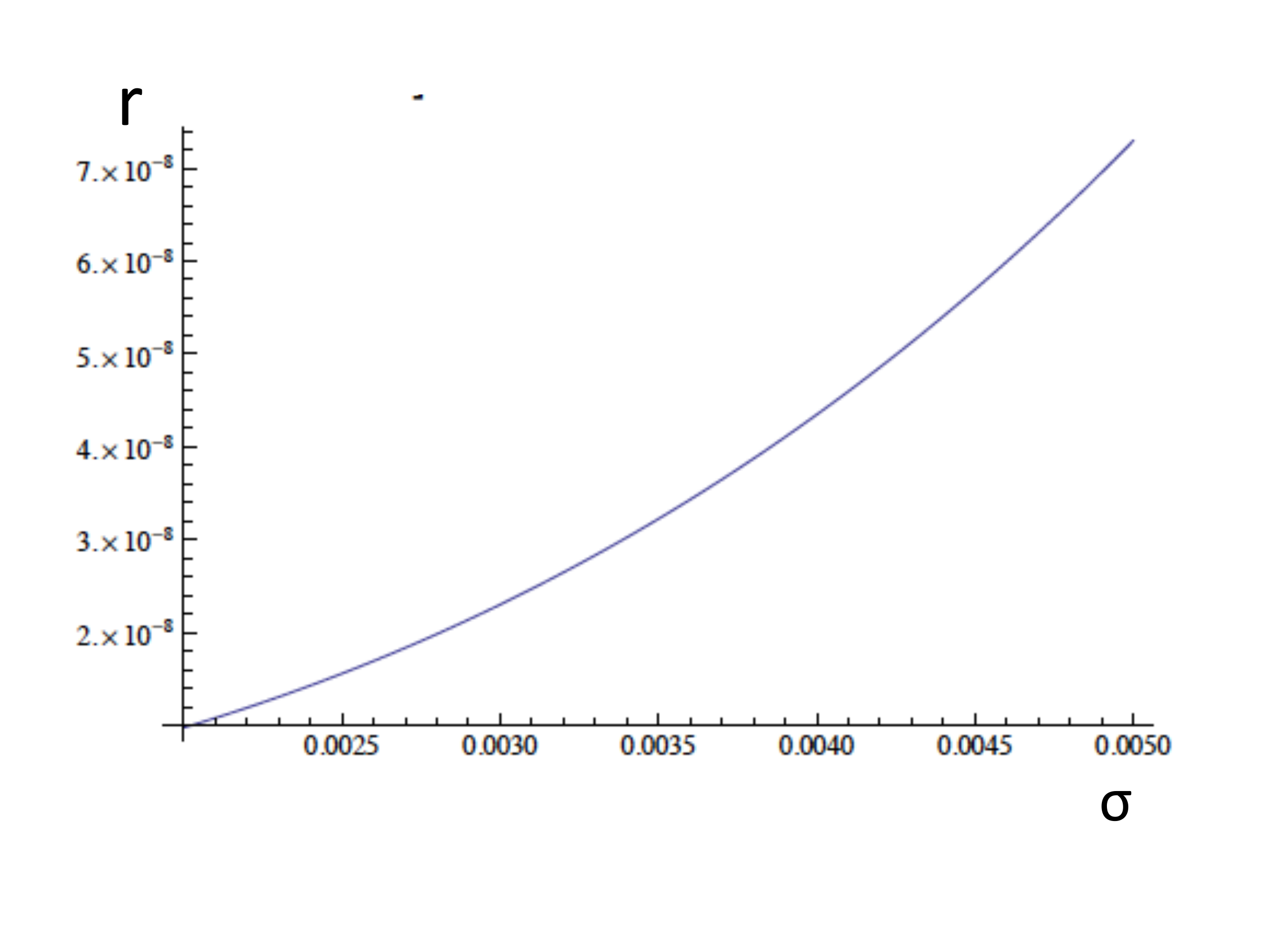} \hfill \includegraphics[width=8cm]{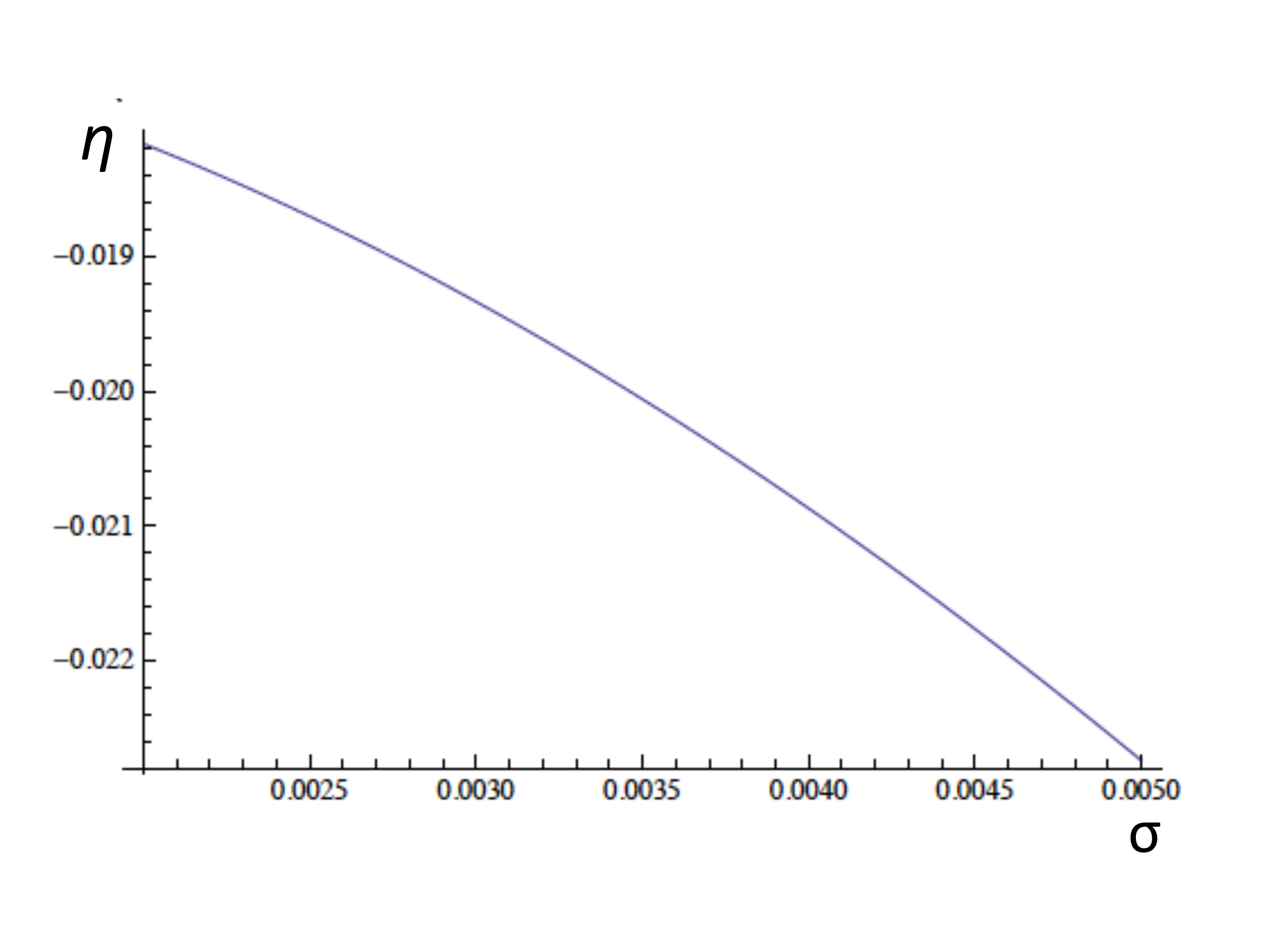} 
 \includegraphics[width=8cm]{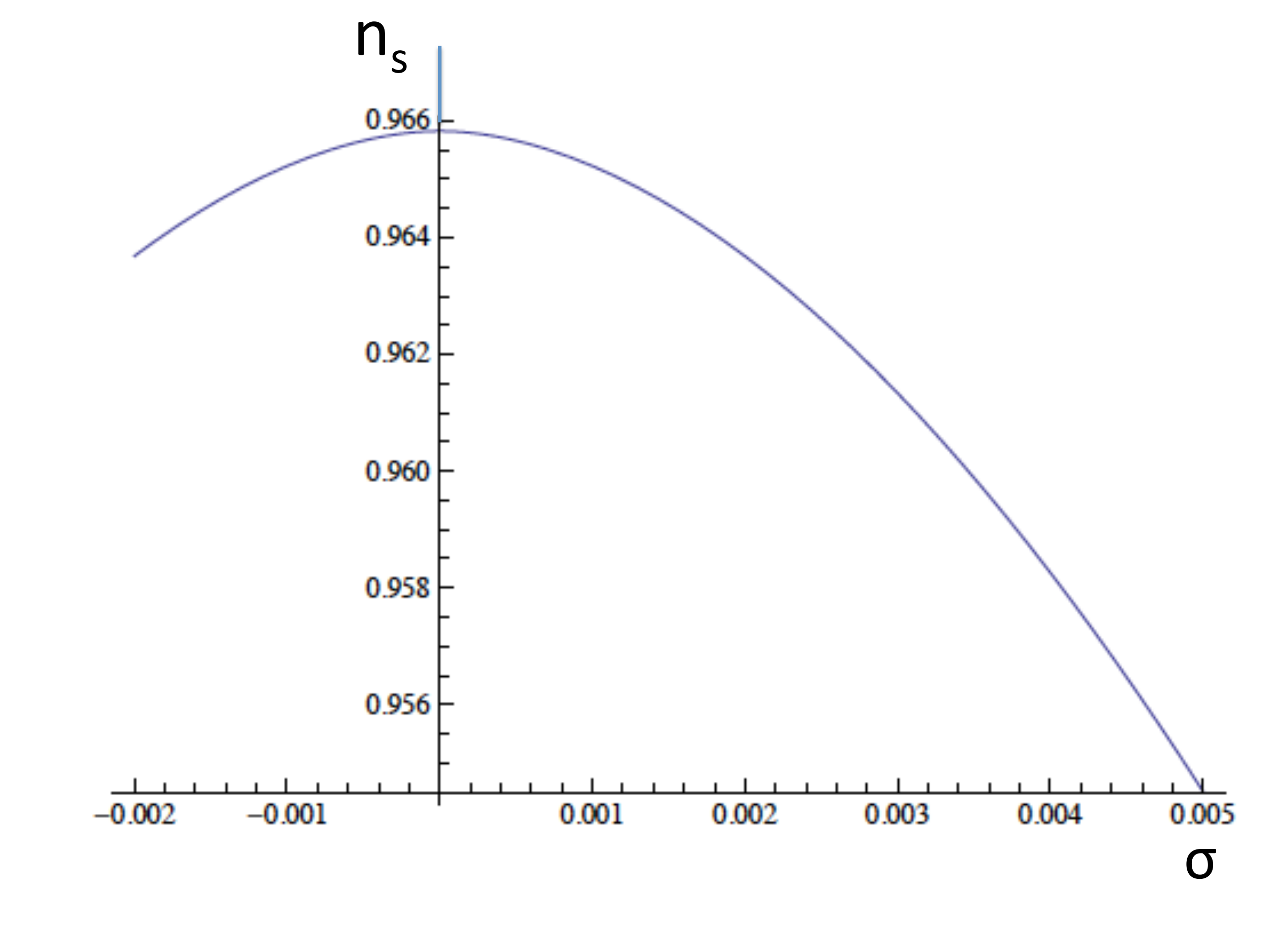} 
\vfill  \includegraphics[width=8cm]{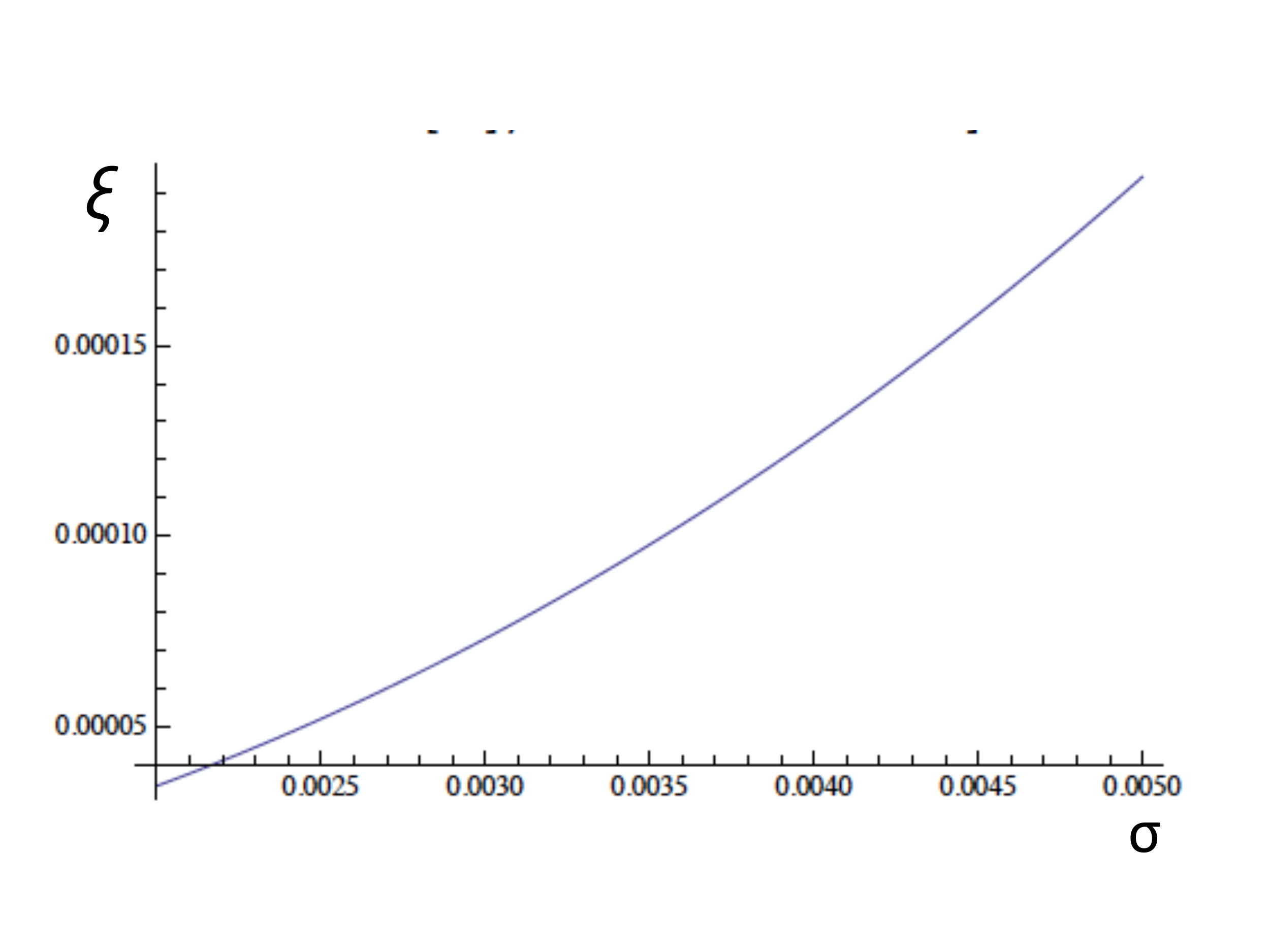} 
\caption{\it The slow-roll parameters (\ref{slowroll}) and the scalar spectral index $n_s$
as functions of the dimensionless inflaton field $\sigma$  for the inflationary scenario induced by 
the one-loop effective potential (\ref{effpotsugra}) of the gravitino condensate field  $\rho(x)$ 
in the effective action (\ref{confsugra2}) of ${\mathcal N}=1$ conformal supergravity. 
Inflation ends well before the gravitino acquires its full
dynamically generated mass  at the non-trivial minimum of the potential at $\rho \sim {\cal O}( {\tilde \kappa}^2)$.}
\label{fig:slow}%
\end{figure}

To ensure a scalar spectral index (\ref{ns}) in agreement with the value measured by Planck~\cite{Planck}, namely
$n_s \simeq 0.960 \pm 0.007$, we require a 
relatively small $\eta$ parameter and thus must choose the dimensionless 
parameter ${\tilde \kappa }^2 \Lambda^2 $ appropriately. 
Typical values we obtain for our scenario are such that the coefficient of the $\sigma^2$ term 
in the effective potential (\ref{effpotsugra}) takes a negative value  $\sim -0.0022$, i.e.:
\begin{equation}\label{kl}
-1 + \frac{11\, {\tilde \kappa}^2 \, \Lambda^2  }{4\, \pi^3} \; = \; -0.0022~,
\end{equation}
while the following value of the cosmological constant $f^2$ guarantees a 
consistent non-trivial minimum of the potential at which the one-loop effective potential vanishes:
\begin{equation}\label{fvalue}
f^2 \; = \; 0.2574\, {\tilde \kappa}^{-4}~.
\end{equation}
With these values the effective potential that we use to analyze inflation acquires the form:
\begin{equation}\label{inflpot}
e^{-1} \, \mathcal{V}_{\rm eff~infl} \; = \; {\tilde \kappa}^{-4}\, \Big[  0.4878 \, \sigma^4  \, 
{\rm ln} \sigma^2  - 0.2549 \, \sigma^4  - 0.0022\, \sigma^2 + 0.2574 \Big]~,
\end{equation}
which is plotted in Fig.~\ref{fig:effpot}. The non-trivial minima of this potential
(which is symmetric about the origin) occur for 
\begin{equation}\label{sigmamin}
\sigma_{\rm min} \; \simeq \; \pm \, 1.0125  \, ,
\end{equation}
implying via (\ref{confsugra2})) a dynamical gravitino mass of order:
\begin{equation}\label{gravinomass2}
 M_{3/2}^{\rm dyn} \; = \; \sqrt{11} \, {\tilde \kappa}^{-1} \, | \sigma_{\rm min}| \,  \simeq 1.0125  \,\sqrt{11}\, \Big(\frac{\kappa}{{\tilde \kappa}} \Big) \, M_{Pl}= 3.3581  \, \Big(\frac{\kappa}{{\tilde \kappa}} \Big) M_{Pl}~,
\end{equation}
which, in view of (\ref{kappalog}), is much lighter than the Planck mass $M_{P} = \sqrt{8 \, \pi} M_{Pl}$,
an important consistency requirement on the model.

As is clear from Fig.~\ref{fig:slow}, the slow-roll inflationary phase occurs for field values in the region 
\begin{equation}\label{inflregion}
0.0025 \pm 0.0005\, < \,  \frac{|\phi |}{M_{Pl}} \, < \, 0.0045 \pm 0.0005~,
\end{equation}
for which the slow-roll parameters (\ref{slowroll}) and the running of the spectral index are well within the 
current experimental constraints~\cite{Planck}.
In particular, during the inflationary slow-roll phase, the (approximately constant) value of the effective potential, 
that fixes the Hubble parameter $H_I$ is  
\begin{equation}\label{effpot3}
\kappa^4 \, V_{\rm eff ~infl} \; \simeq \; \Big(\frac{\kappa}{\tilde \kappa} \Big)^4 \, 0.2574~,  \quad H_I 
\; = \; \frac{\kappa}{\sqrt{3}} V_{\rm eff ~infl}^{1/2} \sim 0.2929 \,  \Big(\frac{\kappa}{\tilde \kappa} \Big)^2 \, M_{Pl} ~.
\end{equation}
As seen in the top panels of Fig.~\ref{fig:slow}, this yields a very small value of $\epsilon = {\mathcal O}(10^{-9})$,
and hence a value of $r = {\mathcal O}(10^{-8} )$ that is far below the Planck sensitivity~\cite{Planck}
and the predictions of the Starobinsky~\cite{starobinski} and Higgs inflation models~\cite{shaposhnikov}. On the other hand,
as seen in the middle panels of Fig.~\ref{fig:slow}, this model yields $\eta  = {\mathcal O}(10^{-2})$ and hence
a Planck-compatible value of $n_s$ for all the values of $\sigma$ displayed. Finally, we see in the bottom
panel of Fig.~\ref{fig:slow} that this model yields $\xi = {\mathcal O}(10^{-4})$,and hence a running spectral index (\ref{alphas}) 
of order $\alpha_s  = {\mathcal O}(10^{-6} - 10^{-5} )$, which is consistent with the current experimental 
data~\cite{encyclopaedia,Planck}.  The number of e-foldings in the region (\ref{inflregion}) is
$N = {\mathcal O}\Big( 30-50 \Big)$. Since there is a single real (composite) inflaton field,
non-Gaussianities in the CMB are negligible, unlike the case of~\cite{alvarez}, where
a complex scalar partner of the Goldstino participates in inflation, leading~\cite{wang} to non-Gaussianities 
linked to the scale of global supersymmetry breaking.

The first part of (\ref{effpot3}) tells us that the magnitude of the potential (\ref{vconstr}) is appropriate for:
\begin{equation}\label{constrsat}
\frac{\tilde \kappa}{\kappa} = {\mathcal O}( 10^3) \gg 1~,
\end{equation}
which is compatible with our assumptions in this work. From (\ref{kappalog}), then, we observe that (\ref{constrsat}) 
is satisfied for transmutation mass scales $\mu$ in the infrared regime:
$ {\rm ln}\Big(\frac{\Lambda^2}{\mu^2}\Big) \, > \, {\mathcal O}(10^8),$ which is a 
consistency check of our analysis. Eq. (\ref{constrsat}) implies a light gravitino with 
a dynamically-generated mass (\ref{gravinomass2}) of order of the GUT scale,
\begin{equation}
\label{m32}
 M_{3/2}^{\rm dyn} \sim 8 \times 10^{15} ~{\rm GeV}~,
\end{equation}
and, on account of (\ref{effpot3}), a Hubble parameter during inflation of order 
\begin{equation}\label{hubbleinfl}
H_I \sim 1.4 \times 10^{-8} \, M_{Pl} \simeq 3.4 \times 10^{10}~{\rm GeV}~, 
\end{equation}
which is well below the upper bound $H_I \, < \, 3.7. \times 10^{-5} \, M_{Pl} ={\mathcal O}( 10^{12})$~GeV 
imposed by the Planck upper limit on the tensor-to-scalar ratio~\cite{Planck}, $ r_{0.002} \, < \, 0.12$ 
(as evaluated at the pivot scale $k_\star = 0.002$~Mpc$^{-1}$). Finally, we obtain from (\ref{fvalue}) a
global supersymmetry-breaking scale of order of the GUT scale, 
\begin{equation}
\sqrt{f} \sim 1.73 \times 10^{15}~{\rm GeV}~.
\end{equation}
At the end of inflation, the gravitino condensate fields rolls fast towards its non-trivial minimum, 
at which local supersymmetry breaks dynamically, with the gravitino acquiring a non-trivial mass (\ref{gravinomass2}). 
The condensate field is massive in this phase, with a squared mass $M_\rho^2$ that is given by the second
derivative of the effective potential with respect to the condensate field, 
evaluated with the value of this field at the non-trivial minimum (\ref{sigmamin}). 
Using (\ref{constrsat}), we find that this mass is of order 
\begin{equation}
\label{condmass}
M_\rho  \sim 2\, \sqrt{2}  \Big(\frac{\kappa}{{\tilde \kappa}}\Big) \, M_{Pl}  = \sqrt{8/11} \, M_{3/2}^{\rm dyn} \sim 7 \times 10^{15} ~{\rm GeV} \sim 0.003 M_{Pl} ~.
\end{equation}
We note that the gravitino-condensate field mass is slightly smaller than the dynamical gravitino mass, 
indicating strong binding energy, unlike the dynamical Higgs case~\cite{bardeen}. However, as we already mentioned,
much more work is needed before definite conclusions on the precise relation between the gravitino 
and condensate masses are reached, which falls beyond the scope of the present work. 
For our purposes here, the above mass estimate should only be viewed as indicative of the order of magnitude.

After inflation, the gravitino condensate performs coherent oscillations about its non-trivial minimum, 
and this phase corresponds to the reheating of the Universe, via decays of the condensate fields to matter and radiation, 
that depends upon the coupling of the supergravity model to matter. We do not discuss any details of
the phenomenology of this phase in the current article, as our purpose here is simply to propose a paradigm
for a novel inflationary scenario due to gravitino condensation that is compatible with the data. 

\section{Summary and Outlook \label{sec:concl}}

We have presented in this paper a new inflationary scenario in supergravity, in which
the inflaton is identified with the gravitino condensate that forms dynamically and breaks
local supersymmetry in a conformal supergravity model, with the complex scalar superfield
corresponding to the conformal factor containing dilaton and axion fields. 
The scale of inflation is linked to the scale $\sqrt{f}$ of global supersymmetry breaking, 
and inflation is associated with an infrared phase, in which the 
gravitino condensate rolls down its potential towards a non-trivial minimum, where the gravitino becomes massive. 
In this way we have obtained a successful model for inflation
that exploits the flatness of the one-loop effective potential of the gravitino condensate near the origin. 
The scenario is a truly minimal inflationary scenario, both in its field content and because
it does not rely on detailed knowledge of the dilaton superfield potential.

We have found the following hierarchy of mass scales in our conformal supergravity model for inflation, 
which is compatible with the current astrophysical data on inflation coming from the Planck satellite~\cite{Planck}:
\begin{eqnarray}
{\rm Planck~Mass}: M_P/{\rm GeV} &=& 1.2 \times 10^{19}\, , \nonumber \\
> \, {\rm gravitino~mass}: M_{3/2}^{\rm dyn}/{\rm GeV}  & \sim & 8 \times 10^{15}  \, , \nonumber \\
> {\rm gravitino}-{\rm condensate~mass}: M_\rho/{\rm GeV}  &=& \sqrt{8/11} M_{3/2}^{\rm dyn}/{\rm GeV} \sim 7 \times 10^{15} \, ,  \nonumber \\ 
> {\rm global~supersymmetry}-{\rm breaking~scale}: \sqrt{f}/{\rm GeV} &=& 1.7 \times 10^{15} \, , \nonumber \\
> \, {\rm Hubble~scale~during~inflation}: H_I/{\rm GeV} & = & 3.4 \times 10^{10} \, , \nonumber \\
{\rm Dilaton~v.e.v.}: e^{-\varphi}  &=& {\mathcal O}\big(10^{3}\big) \, , 
\end{eqnarray}
while the slow-roll inflationary phase occurs for small field values in the region (\ref{inflregion}):
\begin{equation}\label{inflregion2}
6 \times 10^{15} \,  < \,  \frac{|\phi |}{{\rm GeV}} \, < \, 1.1 \times 10^{16} ~, 
\end{equation}
and is characterized by the parameters $\epsilon ={\mathcal O}\big(10^{-9}\big)$, 
$r ={\mathcal O}\big(10^{-8}\big)~, \eta ={\mathcal O}\big(10^{-2}\big)~, \xi ={\mathcal O}\big(10^{-4}\big)$, and 
a Planck-compatible scalar spectral index in the range $n_s \in \big(0.966 - 0.955 \big)$. 

One avenue for further research is the study of gravitino properties such as decays
in such models, along the lines of~\cite{spanos}. The presence of a conformal factor affects the gravitino decay width and,
depending on its value, one may obtain, e.g., in models with neutralino dark matter, a
different density of dark matter relics than in standard supergravity, which could have important phenomenological 
consequences for collider searches of supersymmetry. 

More complete studies in this context should include conformal supergravity models
in which additional scalar fields, such as those appearing in the Standard Model, 
enter the conformal factor as in the analysis of \cite{ferrara2}. 
Such fields have, in the corresponding Jordan frame, non-minimal couplings with the curvature that might lead, 
when combined with our dynamical scenario, to acceptable 
slow-roll conditions for inflation in the early Universe~\cite{shaposhnikov, higgsdil, ferrara2}.  
The success of such a programme would depend on details of the scalar field 
potentials, which induce interactions among those fields and the gravitino condensate in the effective potential. 
These more complicated models would exhibit non-Gaussianities, and it will be interesting to examine their phenomenology in light of Planck satellite data. We do not address such issues here but we hope to come back to them in the future.

\section*{Acknowledgements}

We thank Djuna Croon for discussions. This work was supported in part by the London Centre for
Terauniverse Studies (LCTS), using funding from the European Research
Council via the Advanced Investigator Grant 267352.

\end{document}